\documentclass[a4paper]{article}
\usepackage{RR}
\usepackage{algorithm,algorithmic}
\usepackage{graphicx}
\usepackage{amsmath,amssymb}

\newtheorem{defi}{Definition}

\RRdate{Janvier 2009}

\RRauthor{
	No\"el Malod-Dognin\thanks[irisa]{INRIA Rennes - Bretagne Atlantique, and University of Rennes 1, France.}
  \and
	Rumen Andonov\thanksref{irisa}
  \and
	Nicola Yanev\thanks{Faculty of Mathematics and Informatics, University of Sofia.}\thanks{Institute of Mathematics and Informatics, Bulgarian Academy of Sciences.}
}

\authorhead{N. Malod-Dognin \& R. Andonov \& N. Yanev}

\RRtitle{Probl\`eme de clique maximum\\pour la similarit\'e des structures de prot\'eines}

\RRetitle{Solving Maximum Clique Problem for Protein Structure Similarity}

\RRresume{Une hypoth\`ese fondamentale en biologie mol\'eculaire est que des prot\'eines partageant des structures 3D similaires ont de fortes chances de partager une fonction commune, et dans la plupart des cas d\'erivent d'un anc\^etre commun.
\'Evaluer la similarit\'e entre deux structures de prot\'eines est donc une t\^ache essentielle qui a \'et\'e
largement \'etudi\'ee. 
Comparer deux prot\'eines revient \`a chercher un mariage stable optimal de leurs composants, ce qui est \'equivalent \`a rechercher une clique de poids maximum dans un ``graphe d'alignement" sp\'ecifique.

Dans ce rapport, nous pr\'esentons un nouveau mod\`ele de programmation lin\'eaire en nombres entiers pour r\'esoudre ces probl\`emes de cliques. Ce mod\`ele a \'et\'e impl\'ement\'e en utilisant CPLEX 10. Nous avons \'egalement con\c cu un algorithme de branch and bound d\'edi\'e pour r\'esoudre le probl\`eme de clique de cardinalit\'e maximum. Ces deux approches ont \'et\'e int\'egr\'ees dans VAST (Vector Alignement Search Tool) - un logiciel pour l'alignement des structures 3D de protéines largement utilis\'e au NCBI (National Center for Biotechnology Information). Pour r\'esoudre ses probl\`emes de cliques, VAST utilise l'algorithme de Bron et Kerbosh (BK). Les r\'esultats exp\'erimentaux sur des instances r\'eelles d'alignements de prot\'eines montrent que notre algorithme de branch and bound est jusqu'\`a 116 fois plus rapide que BK pour les plus grandes prot\'eines.
}

\RRabstract{
A basic assumption of molecular biology  is that proteins sharing close three-dimensional (3D) structures are likely to share a common function and in most cases derive from a same ancestor.
Computing the similarity between two protein structures is therefore  a crucial task and has been
extensively investigated.
Evaluating the similarity of two proteins can be done by finding an optimal one-to-one matching between their components, which is equivalent to identifying a maximum weighted clique in a specific ``alignment graph".

In this paper we present a new integer programming formulation  for solving such clique problems. The model has been implemented using the ILOG CPLEX Callable Library. In addition, we designed
a dedicated branch and bound algorithm for solving  the maximum cardinality clique problem. Both approaches have been integrated  in  VAST (Vector Alignment Search Tool) - a software for aligning protein 3D structures largely used in NCBI (National Center for Biotechnology Information). The original  VAST  clique solver uses the well known Bron and Kerbosh algorithm (BK). Our computational results on real life protein alignment instances show that our branch and bound algorithm is up to 116 times faster than BK for the largest proteins.}

\RRmotcle{Branch and bound, programmation lin\'eaire en nombres entiers, clique maximum, alignement de structures de prot\'eines.}

\RRkeyword{Branch and bound, integer programming, maximum clique problems, protein structure alignment.}

\RRprojets{Symbiose}
\RRtheme{\THBio}
\URRennes

\begin{document}
\RRNo{6818} 
\makeRR

\section{Introduction}
A protein is an ordered sequence of amino acids. Under specific physiological conditions, the
linear arrangement of amino acids will fold and adopt a complex 3D shape. This 3D shape contains
some highly regular sub-structures called secondary structures elements (SSEs), such as $\alpha$helices and $\beta$strands.

A fruitful assumption of molecular biology  is that proteins sharing close three-dimensional (3D) structures are likely to share a common function and in most cases derive from a same ancestor.
Computing the similarity between two protein structures is therefore  a crucial task and has been
extensively investigated \cite{1001,xu-gang,halperin,strickland05}.
Evaluating the similarity of two proteins $P_1$ and $P_2$ is usually done by evaluating the best - according to some criterion -  one-to-one matching between their components (also called  alignment).

Among the various protein structure alignment methods, we are interested in {\bf VAST} \cite{gibrat96} (Vector Alignment Search Tool) - a software for aligning protein 3D structures largely used in NCBI (National Center for Biotechnology Information) - where the problem of aligning two proteins is converted into finding a maximum clique in a specific graph.

In VAST, two proteins $P_1$ and $P_2$ are represented by their ordered sets of SSEs ($V_1$ and $V_2$ respectively), and a structural alignment of $P_1$ and $P_2$ is an order-preserving one-to-one matching between the SSEs of $V_1$ and the ones of $V_2$.  Such matchings are represented here by the so called \emph{alignment graph}  $G=(V,E)$.  In our approach,  the vertex set $V$ is depicted by a grid, where each row  corresponds to a SSE of $V_1$, while each column   corresponds  to  a SSE of $V_2$. A vertex $ik$, situated on the  intersection of row $i\in V_1$ with column $k\in V_2$ exists ($ik \in V$),  iff the SSEs $i$ and $k$ are ``compatible" (i.e. are of the same type and have similar lengths). 
An edge $(ik,jl)$ between two vertices $ik$ and $jl$ exists ($(ik,jl) \in E$),  iff  the pair $(ik,jl)$ is ``compatible" (i.e.  $i<j$ and $l<k$ (for order preserving) and if the SSEs couple $i$ and $j$ from $V_1$ can be well superimposed in 3D-space to the  SSEs couple  $k$ and $l$ from $V_2$)\footnote{See \cite{gibrat96} for the exact definition of  superimposing  SSEs couples  in 3D-space.}. To each vertex $ik \in V$ we can associate a weight $S_{ik} \in \mathbb{R}$, and to each edge $(ik,jl) \in E$ we can associate a weight $C_{ikjl} \in \mathbb{R}$. Finding a suitable secondary structure alignment is then equivalent to discovering  a clique in the grid graph  $G$.

Looking for cliques in this kind of graphs arises in different situations, where matching in bipartite graphs preserving the order of vertices is sought. Such an example is another protein structure alignment method called Contact Map Overlap Maximization (CMO) \cite{strickland05}.

Various clique related problems can be formulated in such grid  graph.
The \emph{Maximum Cardinality Clique} problem {\bf MCC} consists in finding in $G$ a clique of maximum cardinality, denoted by MCC($G$). MCC is one of the first problem shown to be NP-Complete \cite{karp72}. 
The \emph{Maximum Vertex Weighted clique} problem {\bf MVW} consists in finding a clique with a maximum sum of vertex weights.  If the vertex weights are all equal to 1, then MVW is equivalent to MCC. Since MCC is a special case of MVW, then MVW is also NP-Complete.
The \emph{Maximum Edge Weighted clique} problem {\bf MEW} consists in finding a clique with a maximum sum of edge weights.
Again, if the weights are all equal to 1, then MEW is equivalent to MCC, so MEW is also NP-Complete.
All these clique problems have been intensively investigated \cite{bron_kerbosh73,abello99,pardalos99,regin03,busygin06}.
Moreover, these three problems are all special cases of the \emph{Maximum Weighted Clique} problem {\bf MWC}, which consists in finding the clique having the maximum sum of vertex and edge weights. If the vertex weights are equal to zero, then MWC is equivalent to MEW, if edge weights are all equal to zero, then MWC is equivalent to MVW. If the vertex weights are all equal to 1 and the edge weights are all equal to zero, then MWC is equivalent to MCC.

In this article, we present a new mathematical programming model for solving the most general clique problem MWC. The  model has been implemented using  ILOG CPLEX 10 Callable Library, and validated on the so called Skolnick set (widely used in protein structure comparison articles \cite{1001,cmos_07,VNS}).  In addition, we also design   a dedicated branch and bound algorithm  (B\&B)  for MCC. 
 The B\&B solver has been integrated in VAST and compared to the original VAST clique solver BK (the Bron and Kerbosh algorithm \cite{bron_kerbosh73}) on large  real life protein structures.  
The obtained results show that our B\&B  algorithm is up to 116 times faster than BK for the largest proteins.

\section{Mathematical programming model for MWC}
The use of mathematical programming is not new in the field of maximum cliques \cite{pardalos92,balas}.
However, by using the properties of our graphs, we designed a new linear mathematical programming model
for solving the maximum weighted clique problem (and thus solving MCC, MVW and MEW) on a grid $G=(V, E)$, where
the weights associated to the vertices and edges are all in $\mathbb{R}$.

To each vertex $ik \in V$ (in row $i \in V_1$ and column $k \in V_2$), we associate a binary variable $x_{ik}$ such that :
\begin{equation*}
    x_{ik} = \left\{
                        \begin{array}{ll}    1 & \text{if vertex } ik \text{ is in the clique}, \\
                                             0 & \text{otherwise.}
                        \end{array} \right.
\end{equation*}
In the same way, to each edge $(ik,jl) \in E$, we associate a binary variable $y_{ikjl}$ such that :
\begin{equation*}
    y_{ikjl} = \left\{
                        \begin{array}{ll}    1 & \text{if edge } (ik,jl) \text{ is in the clique}, \\
                                             0 & \text{otherwise.}
                        \end{array} \right.
\end{equation*}

The goal is to find a clique which maximizes the sum of its vertex weights and the sum of its edge weights. This leads to the objective function :
\begin{equation}\label{obj}
        Z_{MWC} = \max \sum_{ik} S_{ik}~x_{ik} + \sum_{(ik,jl)} C_{ikjl}~y_{ikjl}.
\end{equation}

The one-to-one matching implies that in each row $i \in V_{1}$, at most one vertex can be chosen (only one $x_{ik}$ can be set to 1).
\begin{equation}\label{1b}
    \sum_{k} x_{ik} \leq 1, ~~~\forall i \in V_{1}.
\end{equation}
The same holds for the columns.
\begin{equation}\label{1a}
    \sum_{i} x_{ik} \leq 1, ~~~\forall k \in V_{2}.
\end{equation}
These special order set constraints (maximum one activated vertex per row and per column) lead to compact formulations of the relations between vertices and edges.

Denote by $d^{+}_{col}(ik)$ the set of columns with indices $l$ greater than $k$ and such that there exists at least one edge outgoing from the vertex $ik$ and heading to a vertex in column $l$. In a similar way we introduce the notations  $d^{-}_{col}(ik)$, $d^{+}_{row}(ik)$ and $d^{-}_{row}(ik)$. More precisely,  $d^{-}_{col}(ik)$ is the set of columns with indices $l$ smaller than $k$ and such that there exists at least one edge heading to the vertex $ik$ and coming from a vertex in column $l$. $d^{+}_{row}(ik)$ is the set of rows with indices $j$ greater than $i$ and such that there exists at least one edge outgoing from $ik$ and heading to a vertex in row $j$. And finally, $d^{-}_{row}(ik)$ is the set of rows with indices $j$ smaller than $i$ and such that there exists at least one edge heading to $ik$ and coming from a vertex in row $j$.

Edges driven activations of vertices can be formulated with the following compact inequalities :
    \begin{equation}\label{2a}
        x_{ik} \geq \sum_{j} y_{ikjl}, ~~~\forall ik \in V,~\forall l\in d^{+}_{col}(ik).
    \end{equation}
    \begin{equation}\label{3a}
        x_{jl} \geq \sum_{i} y_{ikjl}, ~~~\forall jl \in V,~\forall k\in d^{-}_{col}(jl).
    \end{equation}
    \begin{equation}\label{2b}
        x_{ik} \geq \sum_{l} y_{ikjl}, ~~~\forall ik \in V,~\forall j\in d^{+}_{row}(ik).
    \end{equation}
    \begin{equation}\label{3b}
        x_{jl} \geq \sum_{k} y_{ikjl}, ~~~\forall jl \in V,~\forall i\in d^{-}_{row}(jl).
    \end{equation}

Vertices driven activations of edges can be formulated with the following compact inequalities :
    \begin{equation}\label{4a}
        \sum_{i} x_{ik} + \sum_{j} x_{jl} - \sum_{ij} y_{ikjl}
                        \leq 1, ~\forall k\in V_2, ~\forall l\in V_2,~k<l.
    \end{equation}
    \begin{equation}\label{4b}
        \sum_{k} x_{ik} + \sum_{l} x_{jl} - \sum_{kl} y_{ikjl}
                    \leq 1, ~\forall i\in V_1, ~\forall j\in V_1,~i<j.
    \end{equation}

\subsubsection*{Remarks: }
Our model is designed to perform on alignment grids, and thus takes advantages of characteristics that grid alike graphs do not to share with randomly generated graphs.
The  properties ``one-to-one matching" and ``order preserving"  lead to the creation of special order sets. In the mathematical model this is  illustrated  by constraints (\ref{1b}) and (\ref{1a}). Moreover, this leads to the implication 
``$(ik,jl)\in E$ implies $i<j$ and $k<l$".  As a consequence, if $ik$ is in a clique $C$, all vertices $jl$ such that $j>i$ and $l<k$, or $j<i$ and $l>k$, cannot be in the clique $C$.

\section{Branch and Bound approach for MCC}

We present here a new branch and bound algorithm (B\&B) for solving the MCC problem in the previously 
defined grid $G=(V,E)$, where the vertices in $V$ are 
labeled by their coordinates $ik$, $i$ for the row
number and $k$ for the column number.

\begin{defi}
A \textbf{successor} of a vertex $ik \in G$ is an element of  the set $\Gamma_G^+(ik) = \{ jl \in V$ s.t. 
$(ik,jl) \in E, i<j $ and $k<l \}$.
\end{defi}

\begin{defi}
A \textbf{predecessor} of a vertex $ik \in G$ is an element of the set $\Gamma_G^-(ik) = \{ jl \in V$ s.t.
$(jl,ik) \in E, j<i $ and $l<k \}$.
\end{defi}

We also use the following notations: 
$\Delta_G^+(ik)$ denotes the subgraph of $G$ induced by the vertices in $\Gamma_G^+(ik)$; ~
$\Delta_G^-(ik)$ denotes the subgraph of $G$ induced by the vertices in $\Gamma_G^-(ik)$; ~
$G^S$ denotes the subgraph of $G$ induced by the set of vertices $S\subset V$.

\subsection{Branch and Bound rules}
\subsubsection{Branching: }\label{branching}

Each node of the B\&B is characterized by a couple ($C$, $Cand$) where $C$ is the clique under construction and $Cand$ is the set of candidate vertices to be added to $C$.
All B\&B nodes can also access to $C_{best}$, the best clique found so far during the exploration of the B\&B tree.
At the root of the B\&B tree, these arguments are initially  set to $C = \emptyset$, $Cand = V$ and $C_{best} = \emptyset$.

For a  given B\&B node $N=(C, Cand)$, the vertices in $Cand$ will be  visited according to their lexicographic increasing order (row first).
We call $NEXT(Cand)$ a function returning the vertex $ik$ in $Cand$ having the smallest lexicographic index.
Denote a direct descendant of $N$ by $N'=(C', Cand')$. 
The first descendant is created by a recursive call with arguments   $C' = C+\{ik\}$ and $Cand'=Cand \bigcap \Gamma_G^+(ik)$. 
This corresponds to exploring deeper the tree staying on the same branch and is realized by the step 7 of algorithm \ref{algo_MCC}. 
Visiting the next direct  descendent of $N$ is done  by a recursive call with arguments 
 $C' = C$ and $Cand'=Cand ~\backslash~ \{ik\}$.  This corresponds  to exploring  a neighboring  branch of the  B\&B  tree  (a wider move) and is realized by the step 8 of algorithm \ref{algo_MCC}.

\subsubsection{Fathoming: }
A leaf in the B\&B tree is interesting only if it contains a clique with a cardinality strictly greater than the cardinality of $C_{best}$.
Being given a B\&B node ($C$, $Cand$), the current best clique $C_{best}$, and a candidate vertex $ik \in Cand$,
denote by $MCC_{ik}(G)$ the maximum cardinality clique in $G$ containing the vertex $ik$.
If $|MCC_{ik}(G)| \leq |C_{best}|$, then we do not miss the solution  by discarding $ik$ from $Cand$ (removing a vertex $ik$ from $Cand$ fathoms all the leafs of the current branch leading to a clique containing $ik$). In our approach we do not compute $|MCC_{ik}(G)|$ but we use some upper bounds (see section \ref{estimator}).

Denote by $C_{ik}$ the best clique that is found by branching on the vertex $ik$, and by $MCC_{ik}(G^{Cand})$ the maximum cardinality clique in $G^{Cand}$ containing $ik$.
It is easily seen that $|C_{ik}| = |C| + |MCC_{ik}(G^{Cand})|$.
Any vertex $ik\in Cand$ such that $|MCC_{ik}(G^{Cand})| \leq |C_{best}| - |C|$ leads to non-interesting leafs, and thus, can be removed from $Cand$.
Again, we are going to use some upper bounds of $|MCC_{ik}(G^{Cand})|$.

\subsubsection{Main algorithm: }
Denote by $REMOVE(Cand, C, C_{best})$ a procedure which removes from $Cand$ all vertices $ik$ such that : 
(i) $|MCC_{ik}(G)| \leq |C_{best}|$, or : (ii) $|MCC_{ik}(G^{Cand})| \leq |C_{best}| - |C|$, according to some upper bounds of $|MCC_{ik}(G)|$ and $|MCC_{ik}(G^{Cand})|$. 

Algorithm \ref{algo_MCC} gives the global procedure MCC\_BB  for solving $MCC(G)$. 

\begin{algorithm}[!ht]
\caption{MCC\_BB($C, Cand, C_{best}$)
\#  initially called with $C = \emptyset$, $Cand = V$ and $C_{best} = \emptyset$}
\label{algo_MCC}
\begin{algorithmic}[1]
\IF {$|C| > |C_{best}|$}
    \STATE $C_{best} \leftarrow C$ \# Recording the new best clique.
\ENDIF
\STATE REMOVE($Cand, C, |C_{best}|$)
\IF {$Cand \neq \emptyset$}
    \STATE $ik \leftarrow NEXT(Cand)$
    \STATE MCC\_BB($C + \{ik\}, Cand ~\bigcap~ \Gamma_G^+(ik), C_{best}$) \# exploring deeper the $ik$  branch 
    \STATE MCC\_BB($C, Cand ~\backslash~ \{ik\}, C_{best}$) \#  visiting another candidate (a wider move)
\ENDIF
\STATE \textbf{return}
\end{algorithmic}
\end{algorithm}

\subsection{Maximum cardinality estimator}\label{estimator}
The efficiency of the MCC\_BB algorithm greatly depends on the ability of the procedure REMOVE to fathom non-interesting vertices from $Cand$.
In order to do this, we need to tightly estimate $|MCC_{ik}(G)|$ and $|MCC_{ik}(G^{Cand})|$.
These bounds are based on what we will call feasible paths in a grid.

\begin{defi}\label{feasiblepath}
A \textbf{feasible path} in a grid $G=\{V, E\}$ is an ordered sequence ``$i_1k_1$, $i_2k_2$, $\ldots$, $i_tk_t$" of vertices $\in V$, such that $\forall n\in [1,t-1]$, $(i_nk_n,i_{n+1}k_{n+1}) \in E$ and  $i_n < i_{n+1}$, $k_n < k_{n+1}$.
\end{defi}
Denote by $P(G)$ the longest (in terms of vertices) feasible path in $G$.
Note that computing $P(G)$ can be done by dynamic programming in $O(|E|)$ time.

\subsubsection{Estimation of $|MCC_{ik}(G)|$ : }\label{mcc_g}
\begin{figure}
    \begin{center}
        \includegraphics[width=7cm]{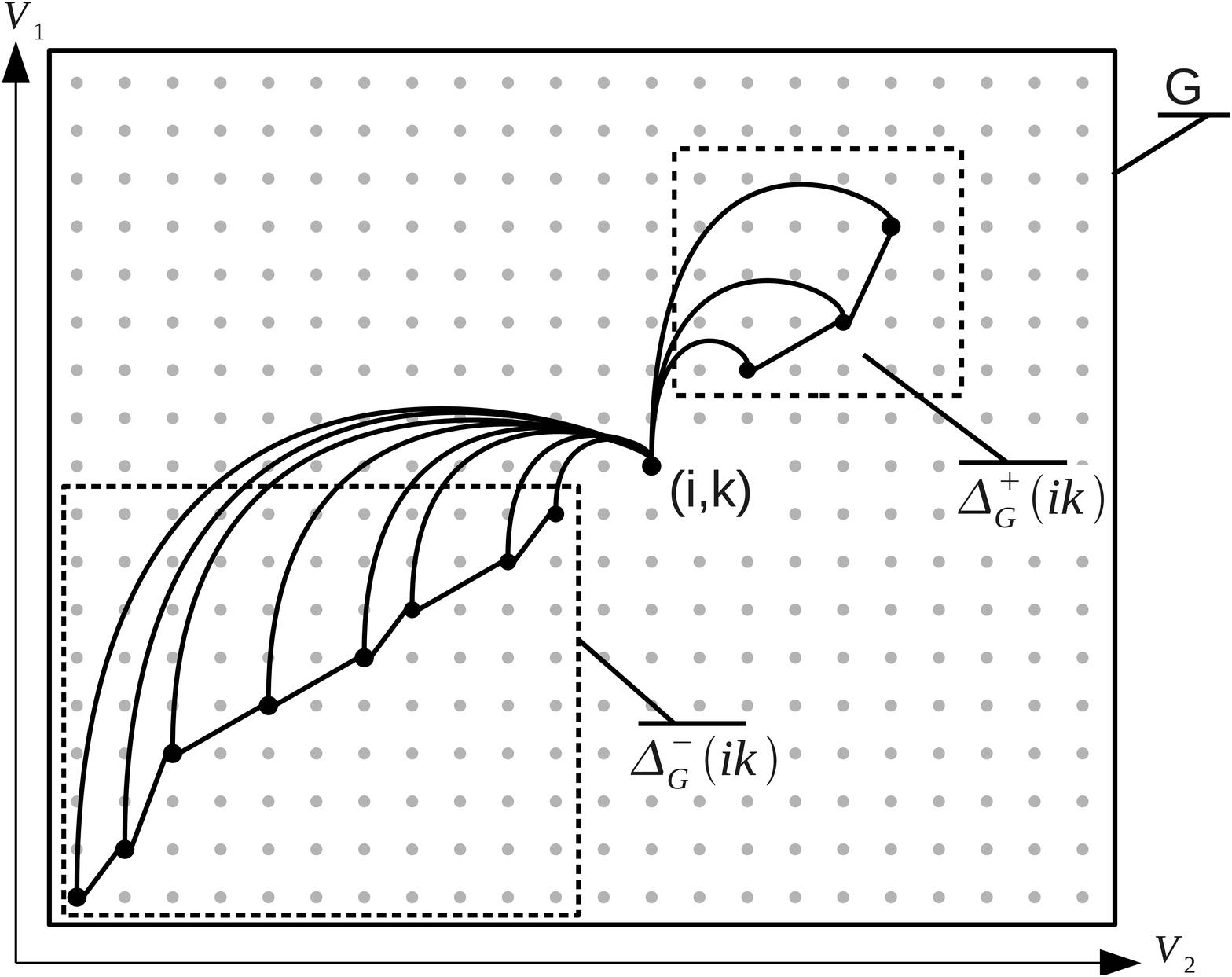}
        \caption{A graphical view of $P_{ik}(G)$. In this example, the longest feasible path in $\Delta^-_G(ik)$ contains 8 vertices,
        the longest feasible path in $\Delta^+_G(ik)$ contains 3 vertices, so the longest feasible path in $G$ such that any vertex is connected to $ik$ contains 12 vertices (including $ik$).}
        \label{p_ik_G}
    \end{center}
\end{figure}

For any vertex $ik\in V$, we  denote by $P_{ik}(G)$ the longest feasible path in $G$ containing $ik$, such that for any vertex
$jl \neq ik$ in the feasible path, $jl$ is connected to $ik$ (i.e. $(ik,jl) \in E$ or $(jl,ik) \in E$).
As illustrated in figure \ref{p_ik_G}, $P_{ik}(G) = P(\Delta^-_G(ik)) \bigcup \{ik\} \bigcup P(\Delta^+_G(ik))$,
and $|P_{ik}(G)| = |P(\Delta^-_G(ik))| + 1 + |P(\Delta^+_G(ik))|$.
It is easily seen that :
\begin{equation}
|MCC_{ik}(G)| \leq |P_{ik}(G)|, \forall ik \in V.
\end{equation}
Thus, any vertex $ik \in Cand$ such that $|P_{ik}(G)| \leq |C_{best}|$ can  be safely removed from
$Cand$. Note that computing all $|P_{ik}(G)|$ can be  done once-for-all in a preprocessing step in  O($|V| \times |E|$) time.

\subsubsection{Estimation of $|MCC_{ik}(G^{Cand})|$ : }\label{mcc_gcand1}
It is obvious that $|MCC(G^{Cand})| \leq |Cand|$, so if $|Cand| \leq |C_{best}| - |C|$ we can safely fathom  all vertices from $Cand$.

\begin{figure}
    \begin{center}
        \includegraphics[width=7cm]{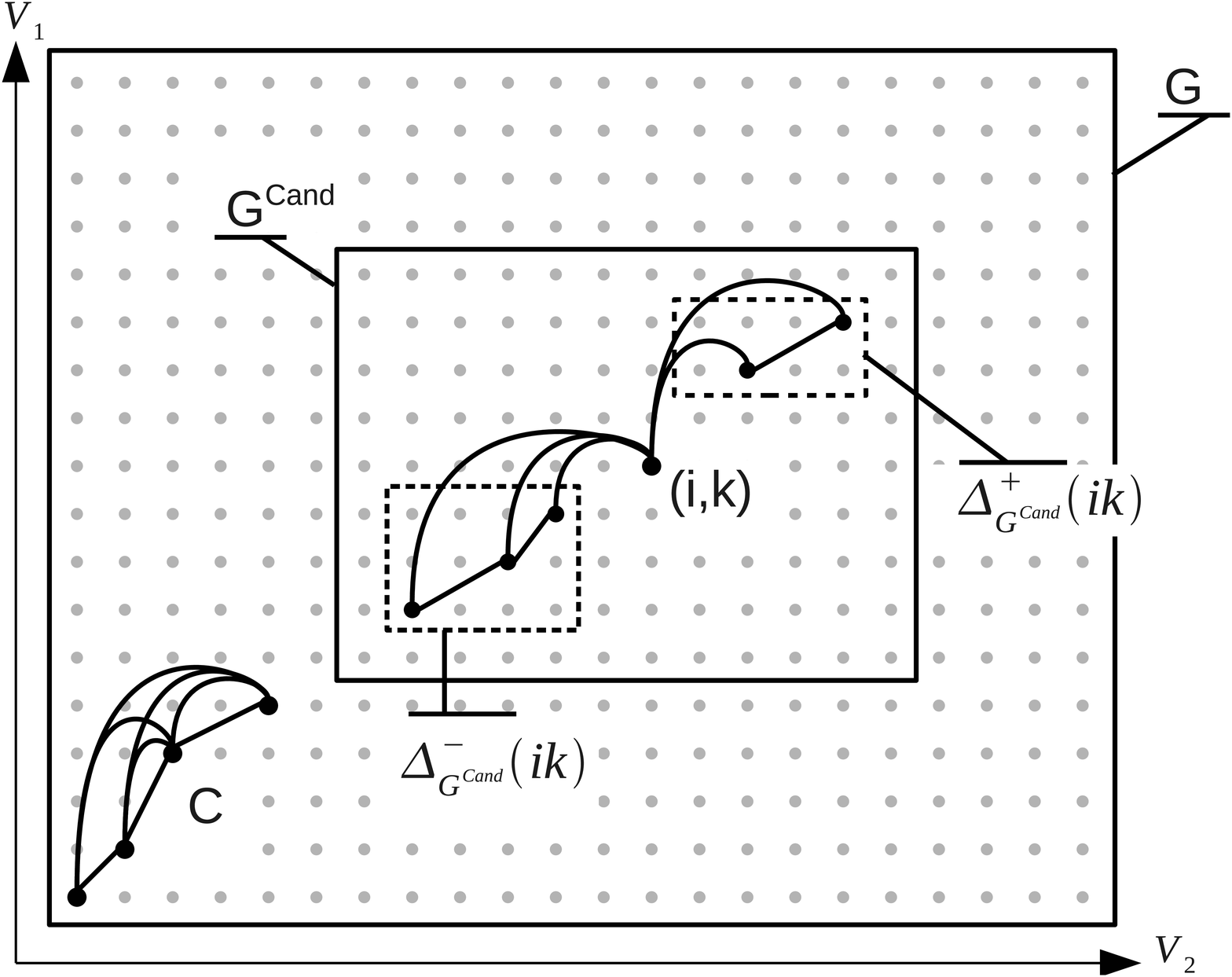}
        \caption{A graphical view of $P_{ik}(G^{Cand})$. In this example, by fixing the clique $C$, we created the
        subgraph $G^{Cand}$ of the candidate vertices to be added to $C$. In this subgraph, for each candidate vertex $ik$,
         $P_{ik}(G^{Cand})$ is found by computing the longest feasible paths in $\Delta^-_{G^{Cand}}(ik)$ and in $\Delta^+_{G^{Cand}}(ik)$.
        }
        \label{p_ik_Gcand}
    \end{center}
\end{figure}

In the same way as we estimated $|MCC_{ik}(G)|$, it is easily seen that :
\begin{equation}
|MCC_{ik}(G^{Cand})| \leq |P_{ik}(G^{Cand})|, \forall ik\in Cand.
\end{equation}
Any vertex $ik \in Cand$ such that $|P_{ik}(G^{Cand})| \leq |C_{best}| - |C|$ can  be safely removed from
$Cand$.
In the subgraph $G^{Cand} = (Cand, E_{Cand})$, computing all $|P_{ik}(G^{Cand})|$ can be done in O($|Cand| \times |E_{Cand}|$) (see figure \ref{p_ik_Gcand}
for a graphical view of $P_{ik}(G^{Cand})$).

Using the abovementioned estimators, we obtain the fathoming procedure REMOVE described in  algorithm  \ref{algo_REM1}.

\begin{algorithm}[!ht]
\caption{REMOVE($Cand, C, C_{best}$)}
\label{algo_REM1}
\begin{algorithmic}[1]
\FOR{$ik \in Cand$}
    \IF{$P_{ik}(G) \leq |C_{best}|$}
        \STATE $Cand \leftarrow Cand ~ \backslash ~\{ik\}$.
    \ENDIF
\ENDFOR
\IF{ $|Cand| \leq |C_{best}| - |C|$}
    \STATE $Cand \leftarrow \{\emptyset\}$
\ENDIF
\FOR{$ik \in Cand$}
    \IF{$P_{ik}(G^{Cand}) \leq |C_{best}| - |C|$}
        \STATE $Cand \leftarrow Cand ~ \backslash ~\{ik\}$.
    \ENDIF
\ENDFOR
\STATE \textbf{return}
\end{algorithmic}
\end{algorithm}

\section{Results}
All results presented here were obtained on the same desktop PC with one Intel Pentium $4^{tm}$ CPU at 3Ghz and 2GB of RAM.
The mathematical programming based solver (MIP), was implemented in C using Ilog Cplex 10.0 Callable Library.
The B\&B solver, 
 was also implemented in C.  This two clique solvers were compared to the original  VAST solver which is based on  Bron and Kerbosh algorithm  (BK) \cite{bron_kerbosh73}. All algorithms were used to solve maximum cardinality clique problems. Note that in fact BK computes and evaluates all maximal cliques in a graph, and hence, can be used to solve any kind of clique problems.

The comparison of the above algorithms  was performed on  real-life  proteins. 
We used two different benchmarks which significantly  differ by the size of the instances (number of SSEs).
The first benchmark is the Skolnick set which was recently largely used in protein structure comparison papers \cite{1001,cmos_07,VNS}). This set  contains 40 small protein chains (containing one domain), with a SSEs number  varying from 5 to 20. The second set benchmark (S2) contains 36 long protein chains (containing from 4 to 6 domains), with a number of SSEs varying from 51 to 87.
Note that for the skolnick set, we only considered instances leading to an alignement graph with at least 100 vertices.

The characteristics of the corresponding alignment grids are given  in table \ref{grids_description}. One peculiarity  is their low density, less than 20\% for the Skolnick set, and less than 6\% for S2 set.
\begin{table}[!ht]
    \begin{center}
        \footnotesize
        \begin{tabular}{|c|c|c|c|c|c|c|c|c|c|}
            \hline
                        & \multicolumn{3}{|c|}{Number of vertices}  & \multicolumn{3}{|c|}{Number of edges} &  \multicolumn{3}{|c|}{Density}    \\
            Set name    & Min   & Average   & Max                   & Min   & Average   & Max               & Min   & Average   & Max           \\
            \hline
            Skolnick    & 100   & 158.92    & 208                   & 886   & 2368.69   & 3547              & 0.16  & 0.18      & 0.20          \\
            \hline
            S2          & 1390  & 2384.97   & 5582                  & 45278 & 144206.44 & 604793            & 0.03  & 0.05      & 0.06          \\
            \hline
        \end{tabular}
        \vspace{-0.5cm}
        \caption{Characteristics of the  grid graphs corresponding to the considered  benchmarks.}
        \label{grids_description}
    \end{center}
    %\vspace{-0.5cm}
\end{table}

\begin{figure}[!ht]
    \begin{center}
    MIP vs BK running time comparison on Skolnick set 
    \includegraphics[angle=270, width = 9cm]{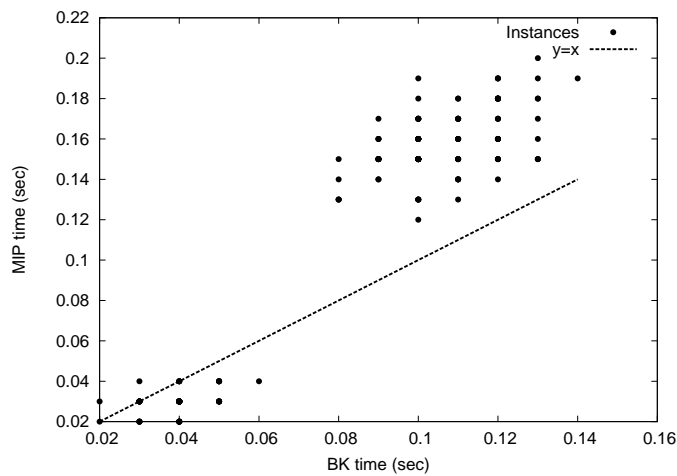}
    \caption{ For each instance the execution time of MIP is plotted on the x-axis, while the one of BK is depicted on the y-axis.  All points are above the $x=y$ line (i.e. BK  is always faster than MIP).}
    \label{skol_bk_mip}
    \end{center}
\end{figure}

Figure \ref{skol_bk_mip} compares the time needed by MIP to the one  of BK on the  Skolnick set.
On the 170 instances containing more than 100 vertices,
MIP is always slower than BK (3.35 times slower in average). This is not surprising, since dedicated solvers are expected to be faster than general purpose solvers  (CPLEX in this case). However, this observation  motivated  us to go further in developing  a fast special purpose clique solver.

\begin{figure}[!ht]
    \begin{center}
    B\&B vs BK running time comparison on S2 set.
    \includegraphics[angle=270, width = 9cm]{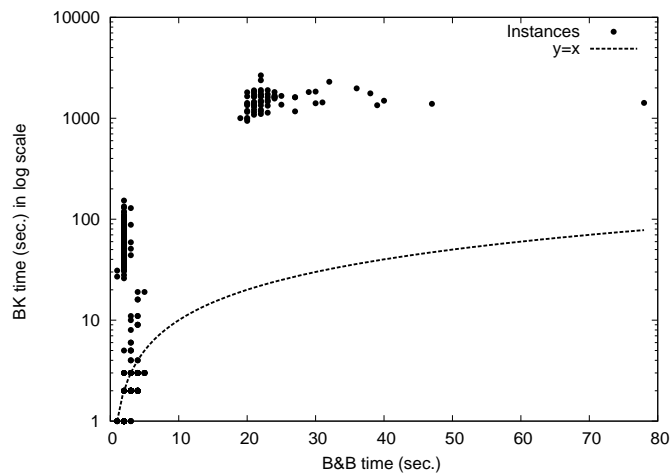}
    \caption{   The execution time of B\&B is presented on the x-axis, while the one of BK is on the y-axis (in log scale).
                The x=y line is also given, and any point above it is an instance for which B\&B is faster than BK.
                In average, B\&B is 15.57 times faster than BK. This superiority  goes up to 116.7 times for the biggest instances.}
    \label{skol_bk_r1}
    \end{center}
\end{figure}

Figure \ref{skol_bk_r1} compares the time needed by B\&B to the one   of BK on set S2.  Here we observed that B\&B is about 15.57 times faster than BK in average,  and on the biggest instances (where both proteins contain more than 80 SSEs), it is up to 116.7 times faster. Such big instances are solved by B\&B in less than 79 seconds (25 sec. on average) while BK needs up to 2660 seconds (1521 sec. on average).
These results are detailed in table \ref{very_large}.

\begin{table}[!ht]
    \begin{center}
        \footnotesize
        \begin{tabular}{|c|c|c|c|c|c|c|c|c|}
            \hline
            Instance            & $|V_1|$ & $|V_2|$ & $|V|$  & $|E|$    & $|MCC|$ & B\&B        & BK \\
                                &         &         &        &          &         & time (sec.) & time (sec.)\\
            \hline
            (d1n6fA\_,d1n6eA\_) & 86 & 83 & 5220 & 526586 & 78 & 22,78 & 2659,27 \\
            \hline
            (d1n6eA\_,d1n6fA\_) & 83 & 86 & 5220 & 527354 & 79 & 22,01 & 2380,55\\
            \hline
            (d1n6fA\_,d1n6dA\_) & 86 & 85 & 5305 & 541073 & 75 & 32,03 & 2296,89\\
            \hline
            (d1n6dA\_,d1n6fA\_) & 85 & 86 & 5305 & 548087 & 75 & 36,81 & 1978,24\\
            \hline
            (d1n6eK\_,d1n6fD\_) & 87 & 87 & 5582 & 604793 & 81 & 22,22 & 1903,39\\
            \hline
            \hline
            (d1n6dD\_,d1n6fF\_) & 85 & 86 & 5304 & 540911 & 75 & 78,91 & 1419,21\\
            \hline
            (d1n6fB\_,d1n6dD\_) & 85 & 85 & 5234 & 523072 & 75 & 47,82 & 1390,55\\
            \hline
            (d1k32F\_,d1n6eG\_) & 85 & 84 & 5279 & 528476 & 76 & 40,08 & 1491,81\\
            \hline
            (d1n6dD\_,d1n6fB\_) & 85 & 85 & 5234 & 532998 & 75 & 39,04 & 1344,8\\
            \hline
            (d1n6dC\_,d1n6fD\_) & 85 & 87 & 5376 & 576604 & 76 & 38,27 & 1765,77\\
            \hline
        \end{tabular}
        \vspace{-0.5cm}
        \caption{Details of S2 benchmark. The density is 0.04 for each instance.
	The first five instances correspond to the biggest running times of BK versus B\&B, while the second five instances
        present the biggest running times of B\&B versus BK.
        Note that aligning $P_1$ with $P_2$ is not the same as aligning $P_2$ with $P_1$, since the superimposition function
        (from \cite{gibrat96}) that we use to define edges is not symmetrical.}
        \label{very_large}
    \end{center}
    %\vspace{-0.5cm}
\end{table}

\section{Conclusion}
We presented a new mathematical programming model for solving the maximum weighted clique problem arising in the context of protein structure comparison.
This model was implemented and validated on a small benchmark.
We also presented a new dedicated branch and bound algorithm for the maximum cardinality clique problem.
The computationnal results show that on big instances, our branch and bound is significantly faster than the Bron and Kerbosh algorithm
(up to 116 times for the largest proteins).
In the near futur, we intend to study the behavior of the proposed algorithms on arbitrary graphs, conveniently transformed into grid graphs
 in a preprocessing step.

\bibliographystyle{unsrt}
\bibliography{biblio}

\end{document}